 \documentclass[preprint ]{aastex} % preprint   (1-column, single-spaced)

\usepackage{emulateapj5}
\usepackage{graphicx}
\usepackage{times}

\makeatletter
\newenvironment{inlinefigure}{%
\def\@captype{figure}%
\noindent\begin{minipage}{0.999\linewidth}\begin{center}}
{\end{center}\end{minipage}\smallskip}
\makeatother

\shorttitle{Reprocessing of Soft X-ray Emission Lines}
\shortauthors{Mauche et al.}

 \slugcomment{Accepted for publication in {\it ApJ\/}         2004 January 13}
\newcommand{\Mdot}{\dot{M}}
\newcommand{\Msun}{\rm M_{\odot}}
\newcommand{\mcg}{MCG $-$6-30-15}

\begin{document}

\title{Reprocessing of Soft X-ray Emission Lines in Black Hole Accretion Disks}

\author{Christopher W.\ Mauche, Duane A.\ Liedahl, Benjamin F.\ Mathiesen}
 \affil{Lawrence Livermore National Laboratory,
       7000 East Avenue, Livermore, CA 94550; \\
       mauche@cygnus.llnl.gov, liedahl1@llnl.gov, mathiesen2@llnl.gov}
\author{Mario A.\ Jimenez-Garate}
 \affil{Massachusetts Institute of Technology, Center for Space Research,
       70 Vassar Street, Cambridge, MA 02139; mario@space.mit.edu} 
\author{\and John C.\ Raymond}
 \affil{Harvard-Smithsonian Center for Astrophysics, 60 Garden Street,
       Cambridge, MA 02138; jraymond@cfa.harvard.edu}

% Abstract:
%---------------------------------------------------------

\begin{abstract}
By means of a Monte Carlo code that accounts for Compton scattering and
photoabsorption followed by recombination, we have investigated the radiation
transfer of Ly$\alpha $, He$\alpha $, and recombination continua photons of
H- and He-like C, N, O, and Ne produced in the photoionized atmosphere of a
relativistic black hole accretion disk. We find that photoelectric opacity
causes significant attenuation of photons with energies above the \ion{O}{8}
K-edge; that the conversion efficiencies of these photons into lower-energy
lines and recombination continua are high; and that accounting for this
reprocessing significantly (by factors of 21\% to 105\%) increases the flux
of the Ly$\alpha $ and He$\alpha $ emission lines of H- and He-like C and O
escaping the disk atmosphere.
\end{abstract}


% Body of the paper:
%---------------------------------------------------------

\section{Introduction}

A controversy currently exists concerning the interpretation of the broad
spectral features observed in the soft X-ray spectra of type 1 Seyfert
galaxies. \citet{lee01} and \citet{tur03} interprete the features seen in
the {\it Chandra\/} High Energy Transmission Grating (HETG) and {\it
XMM-Newton\/} Reflection Grating Spectrometer (RGS) spectra of \mcg\ as
absorption edges produced by a dusty, partially ionized (``warm'') absorber,
while \citet{bra01}, \cite{mas03}, and \citet{sak03b} interpret the
features seen in the {\it XMM-Newton\/} RGS spectra of \mcg\ and Mrk~766 as
relativistically broadened and gravitationally redshifted Ly$\alpha$ emission
lines of H-like C, N, and O arising in the accretion disk surrounding a
maximally spinning Kerr black hole. While warm absorbers are a common feature
of Seyfert 1 galaxies, the relativistic line interpretation is appealing
because it produces model parameters (disk inclination, emissivity index, and
inner radius) that are not inconsistent with those derived for the Fe K$\alpha
$ line.

To be consistent with observations, proponents of the relativistic line
interpretation must explain the apparent absence of the Ly$\beta $ lines and
radiative recombination continua (RRCs) of C, N, and O, as well as emission
by higher-$Z$ ions. \citet{bra01} and \citet{sak03b} propose that this is
accomplished by radiation transfer effects, specifically that the higher-order
Lyman lines are suppressed by line opacity, while the C, N, and O RRCs and
the higher-$Z$ emission features are suppressed by photoelectric opacity.
The production of C, N, and O emission lines in photoionized constant-density
and hydrostatic disk models has been studied quantitatively by \citet{bal02}.
These calculations do not, however, account for the reprocessing of soft
X-ray photons into lower-energy lines and RRCs. In a moderately ionized solar
abundance plasma, the continuum opacity is dominated by C from 0.39 to 0.74
keV and O from 0.74 to 2.4 keV (Fig.~1). If, for example, the medium is
optically thick to photoelectric absorption by O$^{7+}$, a fraction of the
photons with energies greater than that of the \ion{O}{8} K-edge will be
absorbed, and a fraction of those will be reradiated as \ion{O}{8} Ly$\alpha $
photons, thereby increasing the equivalent width of the line. To explore such
processes quantitatively, we have combined a disk atmosphere code with a
Monte Carlo radiation transfer code to calculate the emergent spectrum of
the irradiated atmosphere of a relativistic black hole accretion disk. In this
first communication, we concentrate on the physics of the radiation transfer
of Ly$\alpha $, He$\alpha $, and recombination continua photons of H- and
He-like C, N, O, and Ne, leaving to a future work the calculation of equivalent
widths and detailed comparisons with data.

\section{Disk atmosphere and radiation transfer codes}

The disk atmosphere code calculates the structure of the atmosphere of a black
hole accretion disk photoionized by an external source of nonthermal X-rays.
The code is adapted from that used by \citet{jim02} to model the accretion
disk atmospheres of X-ray binaries. It assumes axial symmetry and solves the
hydrostatic balance, photoionization equilibrium, and radiation transfer
equations on an adaptive mesh in the slab approximation in a succession of
decoupled disk annuli using the \citet{ray93} photoionization plasma code,
which includes the X-ray emissivities and opacities of all ions of the 12
most abundant elements. The disk is assumed to have solar abundances, to be 
gas-pressure dominated, and to have a structure given by \citet{sha73} with
the scale radius $r_\ast = r_g = GM/c^2$, modified by the \citet{pag74} and
\citet{rif95} general relativistic corrections to the vertical structure. We
assume that the incident angle of the nonthermal radiation $i=55^\circ $, to
simulate isotropic illumination, and that the ratio of the incident nonthermal
flux to the disk thermal flux $f= \int F_X d\nu /\int F_{\rm disk} d\nu =
1$. The lower boundary conditions of the calculation are the disk temperature
[$T_{\rm disk}(r)\propto r^{-3/4}$] and pressure, while the upper boundary
condition is the Compton temperature, which is reduced to $T_C\sim 10^7$~K
by the thermal disk emission that irradiates the atmosphere from below. 

As with LMXB disks, the structure of the atmosphere of an AGN disk is sensitive
to the assumptions about the behavior of photoionized plasma on the unstable
branches of the thermal equilibrium S-shaped curve \citep{kro81}. To maximize
the production of soft X-ray lines from H-like ions, we assume thermal
equilibrium solutions on the upper branch of the S-shaped curve. Furthermore,
we ignore the effects of the radiation pressure in the hydrostatic balance
equation, since it reduces the thickness of the atmosphere and the production
of soft X-ray lines, and since in a real disk dynamical and magnetic pressures
may counteract such a reduction \citep{mil00}. We assume that the nonthermal
continuum has a power-law shape with a photon index $\Gamma = 2.1$ with a
high-energy exponential cutoff $E_{\rm high}=150$ keV and a low-energy
exponential cutoff $E_{\rm low}= kT_{\rm disk}$; the black hole has mass
$M=10^7~\Msun $ and spin $a=0.998$, so the radius of the innermost stable
circular orbit $r_{\rm ISCO} = 1.23\, r_g$ \citep{tho74}; the accretion rate
$\Mdot = 10^{24}~{\rm g~s^{-1}} = 0.2\, \Mdot_{\rm Edd}$; and the viscosity
parameter $\alpha = 0.01$. The model has $499\, \hat z\times 35\, \hat r =
17.5\times 10^3$ cells, with the annular radii $r= 1600\, k^{-2}\, r_g$, with
$k=2$,3,4,\ldots ,36. For additional details about the disk atmosphere model,
see \cite{jim04}.

While the radiation transfer equation used in the disk atmosphere code is
sufficiently accurate to calculate the disk atmosphere {\it structure\/}, it
cannot be used to calculate its {\it spectrum\/}. To solve this problem, we
are developing a Monte Carlo code that accounts for radiation transfer in the
atmosphere of a relativistic black hole accretion disk. On a microphysical
scale, we assume that photons are subject to Compton and photoelectric
opacity. For a photon of energy $E$ scattering off an electron drawn from a
population with an isotropic Maxwellian velocity distribution with temperature
$kT$, the Compton cross section
$$\sigma _C= \sigma _T\, \Bigl[1 - 2\Bigl(\frac{E}{m_ec^2}\Bigr)+\frac{26}{5}
\Bigl(\frac{E}{m_ec^2}\Bigr)^2 \Bigl(1+\frac{kT}{m_ec^2}\Bigr)\Bigr],$$
where $\sigma _T$ is the Thomson cross section, which is valid for $E\ll
m_ec^2$ and $kT\ll m_ec^2$ \citep{ryb79}. For a photon with energy $E$ and
direction vector $\hat p$ scattering off an electron with velocity $\beta c$,
$\gamma =(1-\beta ^2)^{-1/2}$, and direction vector $\hat e$, the scattered
photon energy $E'= \Pi'(0)/c$ and direction vector $\hat p' = \Pi'(1{:}3)/
\Pi'(0)$. In these expressions, the scattered photon 4-momentum $\Pi ' =
R_e^{-1}\Lambda ^{-1} R_\gamma ^{-1} S(\Delta\theta ) R_\gamma \Lambda R_e
\Pi $, where $R_e$ and $R_\gamma $ are the electron and photon rotation
matrices, respectively; $\Lambda $ is the Lorentz transformation matrix;
and $S(\Delta\theta ) = Q R(\Delta\theta )$ is the scattering matrix.
In this expression, $Q=1/[1+\gamma E\, (1-\beta\, \hat e \cdot \hat p)\,
(1-\cos\Delta\theta ) /m_ec^2]$ and the scattering angle $\Delta\theta $ is
given by the Klein-Nishina formula for the differential Compton scattering
cross section. The photoelectric opacity is calculated at each energy for 446
subshells of 140 ions of the 12 most abundant elements of a solar abundance
plasma using the partial photoionization cross sections of \citet{ver95}.
Following photoabsorption by K-shell ions, we generate RRC and recombination
line cascades in a probabilistic manner using the recombination cascade
calculations described by \citet{sak99}. The shapes of the RRCs are determined
by the functional form of the photoionization cross sections and the local
electron temperature. Recombination emission from L-shell ions is ignored, 
since it is suppressed by resonant Auger destruction \citep{ros96}.

On a macrophysical scale, we assume that the velocity of the gas is restricted
to the azimuthal direction and has a value $v_\phi = v_K/[1+(v_K/c)^3]$,
where $v_K=\sqrt{GM/r}$ is the Keplerian value, as appropriate for a maximally
spinning Kerr black hole \citep{bar72}; that within each cell the gas moves
with a uniform velocity; and that photons travel in straight lines. By imposing
a limit on the velocity difference from one cell to the next, we determine
the radial and azimuthal dimensions of the cells used in the Monte Carlo
calculation. Specifically, we require that $\Delta v < 2\times 10^8~\rm
cm~s^{-1}$ [set by $\Delta v = \Delta\lambda c/\lambda$ with $\lambda=1.78$
\AA \ (\ion{Fe}{26} Ly$\alpha $) and $\Delta\lambda =0.012$~\AA \ (the {\it
Chandra\/} High Energy Grating FWHM)]. Extending the grid from $r=1.23\, r_g$
to $600\, r_g$, the model has $499\, \hat z \times 108\, \hat r \times
52$--$1005\, \hat\phi = 21.5\times 10^6$ cells.

Photons are propagated in this grid until they are destroyed or escape. From
the ``event list'' of escaping photons, we can calculate the emergent spectrum
in the fluid frame, in the local inertial frame (accounting for Lorentz
boosts), and in the inertial frame at infinity (accounting for gravitational
redshifts). However, because we do not yet follow photons along geodesics,
we are not yet able to directly calculate the observed spectrum at infinity
at a given disk inclination. To solve this problem, we extracted from the
\citet{lao91} model in XSPEC the line profiles $L_k(\nu )$ and relative weights
$w_k$ for a $\delta $-function spectrum for each disk annulus for an inclination angle $i=35^\circ $, as typical of
Seyfert 1 galaxies. The observed spectrum at infinity at this inclination in
the Kerr metric is then $S'(\nu _i) = \sum _k w_k \sum _j S_k(\nu _j)
L_k(\nu _i,\nu _j)$, where $S_k(\nu )$ are the spectra in the fluid frame
for each disk annulus.

%------------------------------------------------------------------------
\begin{inlinefigure}
\centerline{\includegraphics[width=2.50in]{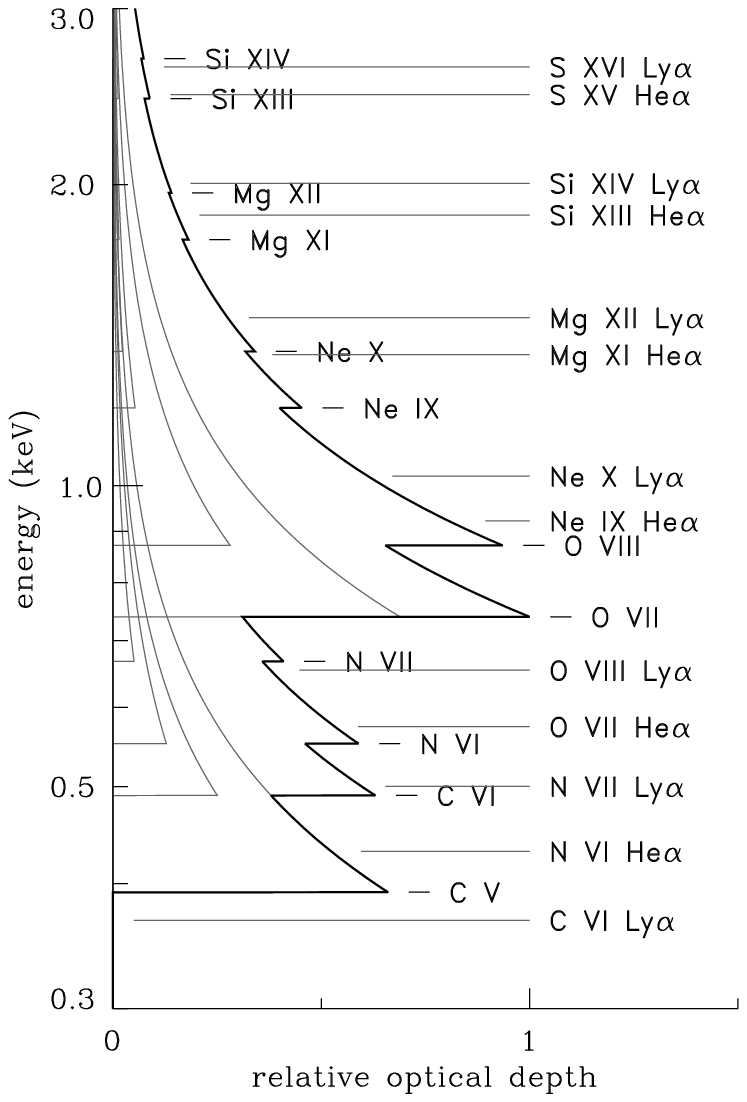}}
\figurenum{1}
\caption{Energy dependence of the optical depth to photoelectric absorption
of a solar abundance plasma with unit ionization fractions of H- and He-like
C, N, O, Ne, Mg, and Si. Energies of the Ly$\alpha $ (He$\alpha $) lines of
H-like (He-like) C, N, O, Ne, Mg, Si, and S are indicated. Note that C (O)
dominates the opacity from 0.39 to 0.74 (0.74 to 2.4) keV.}
\end{inlinefigure}
%------------------------------------------------------------------------

\section{Monte Carlo spectra}

The Monte Carlo code can be configured to run in a number of ways, but for
the present investigation we used it to study the radiation transfer of
recombination photons produced in the disk atmosphere ($\tau _T\le 10$). For
each transition, the emissivities $j_{Z,i}=n_e n_p A_Z f_{Z,i+1}\alpha
T^{-\beta }~\rm photons~cm^{-3}~s^{-1}$, where $n_e$ is the electron density,
$n_p$ is the proton density, $A_Z$ is the abundance of element $Z$ relative
to H, $f_{Z,i+1}$ is the ionization fraction of the $i+1$ charge state of
element $Z$, and $\alpha T^{-\beta}$, is the temperature-dependent ``per-ion''
recombination rate coefficient based on power-law fits to the recombination
rates over the temperature range of interest. Because the recombination
luminosity of our disk atmosphere model is dominated by low-$Z$ ions, we
performed Monte Carlo calculations for the $n=2\rightarrow 1$ (Ly$\alpha $
doublets and He$\alpha $ triplets) and RRC transitions of H- and He-like C, N,
O, and Ne. In addition, we performed calculations for the Ly$\alpha $ doublets
of H-like Mg, Si, S, Ar, Ca, and Fe and for the brightest line of the brightest
Fe L-shell ion (\ion{Fe}{17} $\lambda 17.05$ \AA ), but found for the adopted
disk model that they have only a small effect on the net spectrum.

Figure 2 shows the emergent spectra resulting from the production in the disk
of Ly$\alpha $ photons of H-like O and Ne. In the upper panel, 80\% of the
escaping photons are in the \ion{O}{8} Ly$\alpha $ line and 9\% are in the
\ion{C}{6} Ly$\alpha $ line. In the lower panel, 42\% of the escaping photons
are in the \ion{Ne}{10} Ly$\alpha $ line, 26\% are in the \ion{O}{8} Ly$\alpha
$ line, and 5\% are in the \ion{C}{6} Ly$\alpha $ line. The conversion
efficiency of \ion{Ne}{10} Ly$\alpha $ into \ion{O}{8} Ly$\alpha $ is so high
because the former transition lies just above the \ion{O}{8} K-edge (Fig.~1).
In both cases, very little flux escapes in the N recombination lines because
the C and O opacities dominate over the N opacity, and N recombination lines
are subsequently converted into C recombination lines.

%------------------------------------------------------------------------
\begin{inlinefigure}
\centerline{\includegraphics[width=2.50in]{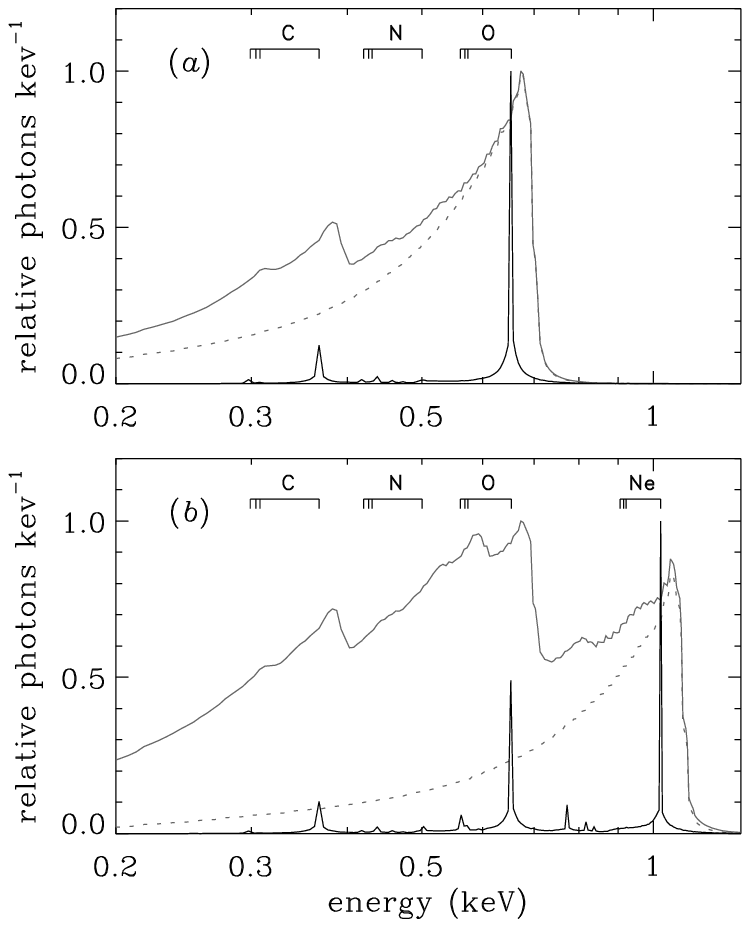}}
\figurenum{2}
\caption{Emergent spectra resulting from the production in the disk of
Ly$\alpha $ photons of H-like O ({\it a\/}) and Ne ({\it b\/}). Black
curves are the spectra in the fluid frame, solid gray curves are the
spectra at infinity at $i=35^\circ $, and the dotted gray curves are
the spectra ignoring recombination following photoionization. Energies of
various Ly$\alpha $ and He$\alpha $ lines are indicated.}
\end{inlinefigure}
%------------------------------------------------------------------------

Figure 3 shows the emergent spectra resulting from the production in the disk
of RRC photons of H-like O and Ne. In the upper panel, 39\% of the escaping 
photons are in the \ion{O}{8} RRC, 30\% are in the \ion{O}{8} Ly$\alpha $ line,
and 6\% are in the \ion{C}{6} Ly$\alpha $ line. In the lower panel, 52\% of
the escaping photons are in the \ion{Ne}{10} RRC, 4\% are in the \ion{Ne}{10}
Ly$\alpha $ line, 16\% are in the \ion{O}{8} Ly$\alpha $ line, and 3\% are in
the \ion{C}{6} Ly$\alpha $ line. The \ion{O}{8} RRC conversion efficiency is
so high because the RRC photons are emitted just above the \ion{O}{8} K-edge,
where the opacity is the highest. The conversion efficiency of \ion{Ne}{10} RRC
photons into the \ion{Ne}{10} Ly$\alpha $ line is so low because the O opacity
dominates over the Ne opacity at the \ion{Ne}{10} K-edge (Fig.~1), and because
\ion{Ne}{10} Ly$\alpha $ photons are subsequently converted into the \ion{O}{8}
Ly$\alpha $ line with high efficiency (Fig.~2{\it b\/}). Once again, the
efficiency of conversion into escaping N recombination lines is very low.

%------------------------------------------------------------------------
\begin{inlinefigure}
\centerline{\includegraphics[width=2.50in]{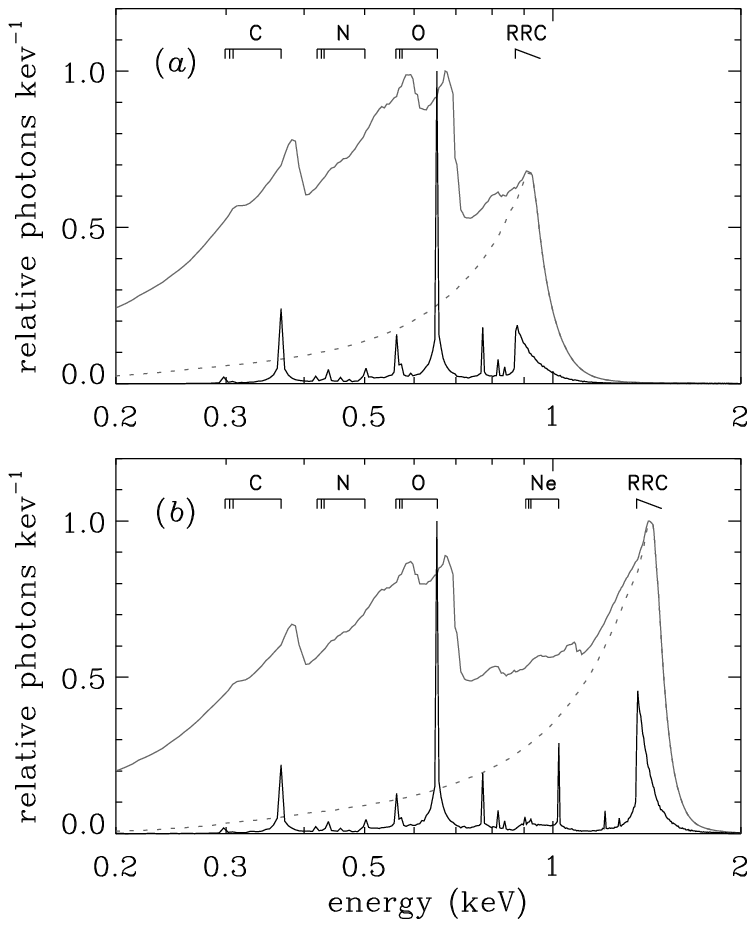}}
\figurenum{3}
\caption{Similar to Figure~2, for the RRC photons of H-like O ({\it a\/}) and
Ne ({\it b\/}). Energies of various Ly$\alpha $ and He$\alpha $ lines and the
RRCs are indicated.}
\end{inlinefigure}
%------------------------------------------------------------------------

As these examples demonstrate, recombination following photoionization
reprocesses soft X-ray photons into lower-energy lines and RRCs, and in the
process increases the strengths and hence the equivalent widths of the C and
O emission features. As shown in Table~1, for the adopted disk model, the
efficiencies of conversion of the Ly$\alpha $, He$\alpha $, and RRC photons
of H- and He-like C, N, O, and Ne into lower-energy lines and RRCs range from
20\% (for \ion{O}{8} Ly$\alpha $) to 86\% (for \ion{Ne}{9} He$\alpha $).
%------------------------------------------------------------------------
\begin{center}
\begin{tabular}{lcccc}
\multicolumn{5}{c}{TABLE 1}\\
\multicolumn{5}{c}{Conversion Efficiencies of Escaping Photons}\\
\hline
\hline
Transition& C& N& O& Ne\\
\hline
\hbox to 1.0in{H-like Ly$\alpha $ \leaders\hbox to 0.4em{\hss.\hss}\hfill}& $\cdots$& 36\%& 20\%& 58\%\\
\hbox to 1.0in{He-like He$\alpha $\leaders\hbox to 0.4em{\hss.\hss}\hfill}& $\cdots$& 30\%& 39\%& 86\%\\
\hbox to 1.0in{H-like RRC         \leaders\hbox to 0.4em{\hss.\hss}\hfill}&     41\%& 54\%& 63\%& 72\%\\
\hbox to 1.0in{He-like RRC        \leaders\hbox to 0.4em{\hss.\hss}\hfill}&     34\%& 33\%& 61\%& 48\%\\
\hline
\end{tabular}
\end{center}
%------------------------------------------------------------------------

To show the net effect of these processes on the observed spectrum, we show in
Figure~4 the full disk atmosphere spectrum accounting for the $n=2\rightarrow
1$ and RRC transitions of H- and He-like C, N, O, and Ne, with relative
normalizations determined by the integrated disk atmosphere emissivities. In
this model, the emission spectrum is dominated by C, N, and O, so the inclusion
of Ne (and higher-$Z$ ions) makes only a minor contribution to the net soft
X-ray spectrum. The emergent spectrum is shown two ways in Figure~4. In the
upper panel, recombination following photoionization is ignored, while in
the lower panel this process is fully accounted for. These spectra differ
in two important ways. First, the lower spectrum is brighter than the upper
spectrum by 50\%: by accounting for recombination following photoionization,
significantly more line flux escapes the disk atmosphere. Specifically,
for the adopted disk model, the flux and hence the equivalent widths of the
\ion{C}{5} He$\alpha $, \ion{C}{6} Ly$\alpha $, \ion{O}{7} He$\alpha $,
and \ion{O}{8} Ly$\alpha $ lines increase by 21\%, 105\%, 28\%, and 44\%,
respectively. Second, the relative line strengths change: the \ion{C}{6}
Ly$\alpha $ to \ion{C}{5} He$\alpha $ line ratio increases by 69\%, the
\ion{O}{8} Ly$\alpha $ to \ion{O}{7} He$\alpha $ line ratio increases by 12\%,
and the summed C/O line ratio increases by 16\%. In both cases, the N emission
features are weak because the N emissivities are relatively low and little
O and Ne flux is reprocessed into N recombination features (note that the
emission feature at 0.5 keV in Fig.~4 is a combination of \ion{N}{7} Ly$\alpha
$ and the \ion{C}{6} RRC).

%------------------------------------------------------------------------
\begin{inlinefigure}
\centerline{\includegraphics[width=2.50in]{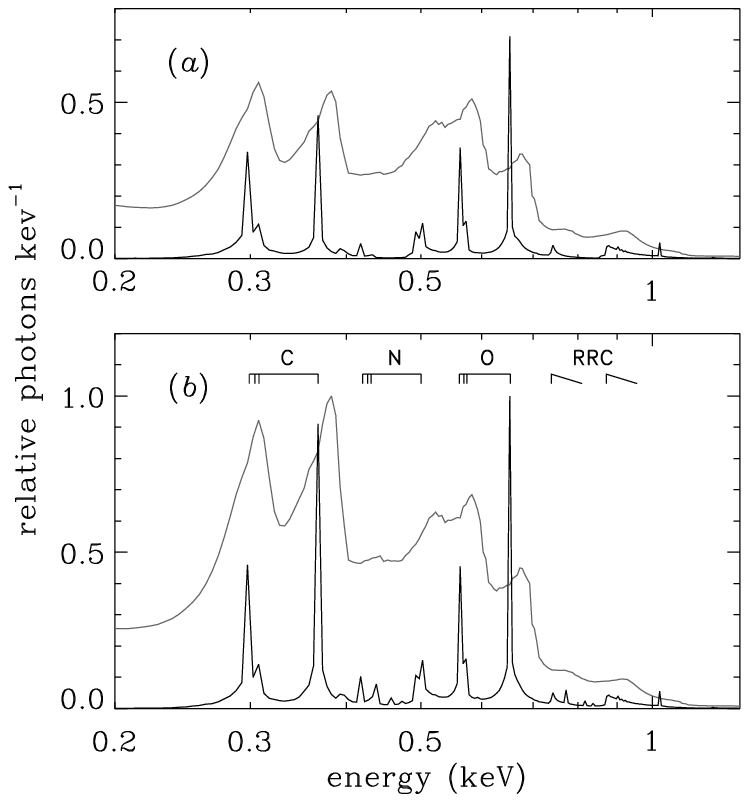}}
\figurenum{4}
\caption{Emergent spectra resulting from the production in the disk of
Ly$\alpha $, He$\alpha $, and RRC photons of H- and He-like C, N, O, and Ne.
Panel ({\it a\/}) ignores recombination emission following photoionization,
while panel ({\it b\/}) accounts for this emission. Black curves are the
spectra in the fluid frame and gray curves are the spectra at infinity at
$i=35^\circ $. Energies of various Ly$\alpha $ and He$\alpha $ lines and the
O RRCs are indicated.}
\end{inlinefigure}
%------------------------------------------------------------------------

\section{Summary and Conclusions}

By means of a Monte Carlo code that accounts for Compton scattering and
photoabsorption followed by recombination, we have investigated the radiation
transfer of Ly$\alpha $, He$\alpha $, and RRC photons of H- and He-like C, N,
O, and Ne produced in the photoionized atmosphere of a relativistic black
hole accretion disk. We find that the conversion efficiencies of these photons
into lower-energy lines and RRCs are high, and that accounting for this
reprocessing significantly (by factors of 21\% to 105\%) increases the flux
of the Ly$\alpha $ and He$\alpha $ emission lines of H- and He-like C and O
escaping the disk atmosphere. These lines dominate the observed spectrum
(Fig.~4{\it b\/}) because the relative emissivities of these ions are high;
the relative photoelectric opacities of these ions are high (Fig.~1), so
photons with energies of 0.39 to 2.4 keV are preferentially absorbed by C and
O and converted into line and RRC photons (Fig.~2); and RRC photons are
converted with high efficiencies into lines (Fig.~3). The relative strength of
the \ion{C}{6} Ly$\alpha $ line is particularly high in our model because it
is fed strongly by all higher-$Z$ ions.

Our model spectrum appears to differ from the {\it Chandra\/} HETG and {\it
XMM-Newton\/} RGS spectra of \mcg \ and Mrk~766 in having too much flux in
the \ion{C}{5} and \ion{O}{7} He$\alpha$ lines and too little flux in the
\ion{N}{7} Ly$\alpha $ line. The relative strength of the \ion{C}{5} and
\ion{O}{7} He$\alpha$ lines may not be of concern, since it should be possible
to weaken these lines by increasing the mean ionization level of the disk
atmosphere: in their photoionized constant-density disk models, BRF find
that the Ly$\alpha $ lines dominate the soft X-ray spectrum for ionization
parameters $\xi\gtrsim 500$. In contrast, the relative weakness of the
\ion{N}{7} Ly$\alpha $ line may be of concern. BRF find that the strength of
this line can be increased simply by increasing the N abundance. However, we
note that if the opacity from 0.55 to 0.67 keV comes to be dominated by $\rm
N^{5+}$, it and not C will capture the \ion{O}{8} Ly$\alpha $ and \ion{O}{7}
He$\alpha $ photons, leading to an increase in the \ion{N}{6} He$\alpha $
line strength. \citet{sak03a} has proposed an alternate way to increase the
\ion{N}{7} Ly$\alpha $ line strength by a Bowen-like fluorescence mechanism,
wherein \ion{O}{8} Ly$\alpha $ photons pump the \ion{N}{7} Ly$\zeta $
transition, and the resulting \ion{N}{7} Ly$\zeta $ photons are converted into
\ion{N}{7} Ly$\alpha $ photons via line opacity. We are not yet able to test
this mechanism quantitatively, but we expect that it will not be particularly
efficient in an irradiated disk atmosphere, since the \ion{O}{8} Ly$\alpha $
photons are produced in a layer where the $\rm O^{7+}$ ionization fraction
is low and the $\rm N^{6+}$ ionization fraction is nearly zero. Only those
\ion{O}{8} Ly$\alpha $ photons that are directed down into the disk {\it
and\/} survive passage through the $\rm O^{7+}$ layer will have a chance of
being converted into \ion{N}{7} Ly$\alpha $ photons.

In its current form, our Monte Carlo code treats in considerable detail many
of the physical processes that affect the transfer of X-ray photons in the
photoionized atmosphere of a relativistic black hole accretion disk, but it
does not yet follow photons along geodesics and it does not yet account for
line opacity. When these processes are included in the code, we will be able
to directly calculate the disk spectrum observed at a given inclination at
infinity and study the Bowen fluorescence mechanism, the radiation transfer of
higher-order Lyman lines, and many other physical processes that affect the
X-ray spectra of AGN disks.

\acknowledgments
The authors acknowledge with gratitude numerous helpful discussions with
Masao Sako and Diego Torres. This work was performed under the auspices of
the U.S.~Department of Energy by University of California Lawrence Livermore
National Laboratory under contract No.~W-7405-Eng-48.

% References
%---------------------------------------------------------


\begin{thebibliography}{}
\bibitem[Ballantyne, Ross, \& Fabian(2002, hereafter BRF)]{bal02} 
         Ballantyne, D.~R., Ross, R.~R., \& Fabian, A.~C.\ 2002, \mnras,
         336, 867, BRF 
\bibitem[Bardeen, Press, \& Teukolsky(1972)]{bar72}
         Bardeen, J.~M., Press, W.~H., \& Teukolsky, S.~A.\ 1972, \apj, 
         178, 347 
\bibitem[Branduardi-Raymont et al.(2001)]{bra01} 
         Branduardi-Raymont, G., Sako, M., Kahn, S.~M., Brinkman, A.~C.,
         Kaastra, J.~S., \& Page, M.~J.\ 2001, \aap, 365, L140
\bibitem[Jimenez-Garate, Raymond, \& Liedahl(2002)]{jim02} 
         Jimenez-Garate, M.~A., Raymond, J.~C., \& Liedahl, D.~A.\ 2002,
        \apj, 581, 1297
\bibitem[Jimenez-Garate et al.(2004)]{jim04} 
         Jimenez-Garate, M.~A., Raymond, J.~C., Liedahl, D.~A., Mauche,
         C.~W.\ 2004, in preparation
\bibitem[Krolik, McKee, \& Tarter(1981)]{kro81}
         Krolik, J.~H., McKee, C.~F., \& Tarter, C.~B.\ 1981, \apj, 249, 422 
\bibitem[Laor(1991)]{lao91}
         Laor, A.\ 1991, \apj, 376, 90 
\bibitem[Lee et al.(2001)]{lee01}
         Lee, J.~C., Ogle, P.~M., Canizares, C.~R., Marshall, H.~L., Schulz,
         N.~S., Morales, R., Fabian,  A.~C., \& Iwasawa, K.\ 2001, \apjl,
         554, L13
\bibitem[Mason et al.(2003)]{mas03}
         Mason, K.~O., et al.\ 2003, \apj, 582, 95 
\bibitem[Miller \& Stone(2000)]{mil00}
         Miller, K.~A., \& Stone, J.~M.\ 2000, \apj, 534, 398 
\bibitem[Page \& Thorne(1974)]{pag74}
         Page, D.~N.~\& Thorne, K.~S.\ 1974, \apj, 191, 499
\bibitem[Raymond(1993)]{ray93}
         Raymond, J.~C.\ 1993, \apj, 412, 267 
\bibitem[Riffert \& Herold(1995)]{rif95}
         Riffert, H., \& Herold, H.\ 1995, \apj, 450, 508
\bibitem[Ross, Fabian, \& Brandt(1996)]{ros96}
         Ross, R.~R., Fabian, A.~C., \& Brandt, W.~N.\ 1996, \mnras,
         278, 1082
\bibitem[Rybicki \& Lightman(1979)]{ryb79}
         Rybicki, G.~B., \& Lightman, A.~P.\ 1979, Radiative Processes in
         Astrophysics (New York: Wiley)
\bibitem[Sako(2003)]{sak03a}
         Sako, M.\ 2003, \apj, 596, 114
\bibitem[Sako et al.(2003)]{sak03b}
         Sako, M., et al.\ 2003, \apjl, 596, 114
\bibitem[Sako et al.(1999)]{sak99}
         Sako, M., Liedahl, D.~A., Kahn, S.~M., \& Paerels, F.\ 1999, \apj,
         525, 921 
\bibitem[Shakura \& Sunyaev(1973)]{sha73}
         Shakura, N.~I, \&  Sunyaev, R.~A.\ 1973, \aap, 24, 337 
\bibitem[Thorne(1974)]{tho74}
         Thorne, K.~S.\ 1974, \apj, 191, 507 
\bibitem[Turner et al.(2003)]{tur03}
         Turner, A.~K., Fabian, A.~C., Vaughan, S., \& Lee, J.~C. 2003,
        \mnras , 346, 833 
\bibitem[Verner \& Yakovlev(1995)]{ver95}
         Verner, D.~A., \& Yakovlev, D.~G.\ 1995, \aaps, 109, 125 

\end{thebibliography}
\end{document}